\documentclass[letter,longauth,traditabstract]{aa} 
\usepackage{epsfig}
\usepackage{natbib}
\usepackage{txfonts}
\begin{document}
\title{Transiting exoplanets from the CoRoT
  space mission}

\subtitle{IV. CoRoT-Exo-4b: A transiting planet in a 9.2 day
  synchronous orbit\thanks{The CoRoT space mission, launched on
    December $27^{\rm th}$ 2006, has been developed and is operated by
    CNES, with the contribution of Austria, Belgium, Brasil, ESA,
    Germany and Spain. The first CoRoT data will be available to the
    public from February 2009 through the CoRoT archive: {\tt
      http://idoc-corot.ias.u-psud.fr/}.}}

   \author{S.\ Aigrain\inst{1} \and
     A. Collier Cameron\inst{2} \and 
     M. Ollivier\inst{3} \and 
     F. Pont\inst{1} \and 
     L. Jorda\inst{4} \and 
     J. M. Almenara\inst{5} 
     R. Alonso\inst{4} \and 
     P. Barge\inst{4} \and 
     P. Bord{\' e}\inst{3} \and 
     F. Bouchy\inst{6} \and 
     H. Deeg\inst{5} \and 
     R. De la Reza\inst{7} \and 
     M. Deleuil\inst{4} \and 
     R. Dvorak\inst{8} \and 
     A. Erikson\inst{9} \and 
     M. Fridlund\inst{10} \and 
     P. Gondoin\inst{10} \and 
     M. Gillon\inst{11} \and 
     T. Guillot\inst{12} \and 
     A. Hatzes\inst{13} \and 
     H. Lammer\inst{14} \and 
     A. F. Lanza\inst{15} 
     A. L{\' e}ger\inst{3} \and 
     A. Llebaria\inst{4} \and 
     P. Magain\inst{16} \and 
     T. Mazeh\inst{17} \and 
     C. Moutou\inst{4} \and 
     M. Paetzold\inst{18} \and 
     C. Pinte\inst{1} \and 
     D. Queloz\inst{11} \and 
     H. Rauer\inst{9,19} \and 
     D. Rouan\inst{20} \and 
     J. Schneider\inst{21} \and 
     G. Wuchter\inst{13} \and 
     S. Zucker\inst{22} 
          }

          \institute{
            School of Physics, University of Exeter, Exeter, 
            EX4 4QL, UK \and
            Sch.\ Physics \& Astronomy, Univ.\ St Andrews,
            St Andrews KY16 9SS, UK \and
            IAS, Universit{\' e} Paris XI, 91405 Orsay, France \and
            LAM, Universit{\' e} de Provence, 13388
            Marseille, France \and
            IAC, E-38205 La Laguna, Spain \and
            IAP, Universit{\' e} Pierre \& Marie Curie,
            75014 Paris, France \and
            ON/MCT, 20921-030, Rio de Janeiro, Brazil \and
            IfA, University of Vienna, 1180 Vienna, Austria \and
            Institute of Planetary Research, DLR, 12489 Berlin,
            Germany \and
            RSSD, ESA/ESTEC, 2200 Noordwijk, The
            Netherlands \and
            Observatoire de Gen{\` e}ve, 1290 Sauverny, Switzerland
            \and
            OCA, CNRS UMR 6202, BP 4229, 06304 Nice Cedex 4, France
            \and
            Th{\" u}ringer Landessternwarte, 07778
            Tautenburg, Germany \and
            IWF, Austrian Academy of Sciences,
            A-8042 Graz, Austria \and
            INAF -- Osservatorio Astrofisico di Catania, 95123
            Catania, Italy \and
            IAG, Universit{\' e} de Li{\` e}ge, Li{\` e}ge 1, Belgium
            \and
            Sch.\ Physics \& Astronomy, Tel Aviv Univ., Tel
            Aviv 69978, Israel \and
            RIU,
            Universit{\"a}t zu K{\" o}ln, 50931 K{\" o}ln, Germany \and
            ZAA, TU Berlin,
            D-10623 Berlin, Germany \and
            LESIA, Observatoire de Paris, 92195 Meudon,
            France \and
            LUTH, Observatoire de Paris, 92195 Meudon, France \and
            Dept.\ Geophysics \& Planetary Sciences, Tel Aviv
            Univ., Tel Aviv 69978, Israel}

   \date{Received \ldots; accepted \ldots}

 
  \abstract
  {CoRoT, the first space-based transit search, provides ultra-high
    precision light curves with continuous time-sampling over periods
    of up to 5 months. This allows the detection of transiting planets
    with relatively long periods, and the simultaneous study of the
    host star's photometric variability. 
    In this letter, we report on the discovery of the transiting giant
    planet CoRoT-Exo-4b and use the CoRoT light curve to perform a
    detailed analysis of the transit and to determine the stellar
    rotation period. 
    The CoRoT light curve was pre-processed to remove outliers and
    correct for orbital residuals and artefacts due to hot pixels on
    the detector. After removing stellar variability around each
    transit, the transit light curve was analysed to determine the
    transit parameters. A discrete auto-correlation function method
    was used to derive the rotation period of the star from the
    out-of-transit light curve. 
    We derive periods for the planet's orbit and star's rotation of
    $9.20205 \pm 0.00037$ and $8.87\pm 1.12$ days respectively,
    consistent with a synchronised system. We also derive the
    inclination, $i=90.00_{-0.085}^{+0.000}$ in degrees, the ratio of
    the orbital distance to the stellar radius, $a/R_{\rm
      s}=17.36_{-0.25}^{+0.05}$, and the planet to star radius ratio
    $R_{\rm p}/R_{\rm s}=0.1047_{-0.0022}^{+0.0041}$. 
    We discuss briefly the coincidence between the orbital period of
    the planet and the stellar rotation period and its possible
    implications for the system's migration and star-planet
    interaction history.}


   \maketitle
%

\section{Introduction}
\label{intro}

Transits provide unique insights into fundamental aspects of
extra-solar planets that are currently beyond the reach of other
techniques: mean density (through the determination of true masses and
radii), atmospheres (through transmission spectroscopy and secondary
transit observations), and formation and evolution mechanisms (through
the statistics of the orbital parameters, including the true
mass). All but two of the transiting planets published so far have
been very short-period planets ($<5$\,d), whose properties are bound
to be affected by extreme proximity to their host star. The exceptions
are HD\,147506b \citep{bkt+07} and HD\,17156b \citep{bal+07},
with periods of 5.6 and 21.2\,d, both in very eccentric orbits ($e \ge
0.5$).

CoRoT is the first space-based transit survey, and over its lifetime
will survey 120\,000 stars for up to 5\,months with precisions down to
0.1\,mmag per hour. This letter is the $4^{\rm th}$ in a series
presenting new extra-solar planets discovered thanks to the CoRoT
observations and associated ground-based follow-up program. Papers I
\citep{bba+08} and II \citep{aab+08} present the discoveries of
CoRoT-Exo-1b and CoRoT-Exo-2b, both very close-in giant planets. Paper
III \citep{bqd+08} reports on the observation of the spectroscopic
transit of CoRoT-Exo-2b. Here we report on the detection of
CoRoT-Exo-4b, a relatively long period (9.2\,d) gas-giant transiting
planet, and on the analysis of the CoRoT light curve (the production
of which is described in Section~\ref{obs}) to determine the transit
parameters (see Section~\ref{transit}) and the stellar rotation period
(see Section~\ref{star}). This analysis makes no assumptions aside
from the planetary nature of the companion, which was confirmed
through ground-based follow-up observations which are reported in a
companion letter (\citealt{mbg+08} hereafter Paper V). In Paper V, the
full system parameters are determined using the transit parameters
reported in the present work. These parameters are used in
Section~\ref{concl}, where we discuss the implications of the
relatively long period and of its proximity to the stellar rotation
period. A further detection, CoRoT-Exo-3b, will be presented in
\citet[][Paper VI]{dha+08}.

\section{Observations and data processing}
\label{obs}

Like the previously published CoRoT planets, CoRoT-Exo-4b was first
detected during near real-time analysis of raw data, the `alarm mode'
(which is described in Papers I and II), triggering the ground-based
follow-up described in Paper V, which included ground-based photometry
in- and out-of-transit, at higher spatial resolution than CoRoT's own,
to identify which of the stars falling in the CoRoT aperture was being
eclipsed, and multiple radial velocity measurements to derive the
companion mass. Once the planetary nature of the companion was
confirmed, a high resolution, high signal-to-noise spectrum of the
host star was obtained to derive accurate stellar parameters.

CoRoT-Exo-4 (GSC designation $0480002187$), whose coordinates and
magnitude are given in Table~\ref{lcpar}, was observed as part of
CoRoT's initial run, during which $\sim 12\,000$ stars with magnitude
$12<R<16$ falling in a $1.3^{\circ} \times 2.6^{\circ}$ pointing close
to the anticentre of the Galaxy were monitored nearly continuously for
58 days, starting on the $6^{\rm th}$ of February 2007. A total of
$72319$ flux measurements were obtained for CoRoT-Exo-4. For the first
33\,d of the run, the time sampling is 512\,s, after which it was
switched to 32\,s as the transits were detected by the alarm mode.

Aperture photometry through a mask, automatically selected from a set
of 256 templates at the beginning of the run, is performed on
board. For stars brighter than $R=15$, the flux is split along
detector column boundaries into broad-band red, green and blue
channels. Although the transits were detected in the raw data, the
analysis presented here was based on the pipeline-processed light
curve. The pipeline \citep{abb+08} currently includes background
subtraction and partial jitter correction. For each exposure, a global
background level is estimated from a handful of $10\times 10$ pixel
background windows distributed over each CCD, excluding background
windows affected by hot pixels, and subtracted. The jitter correction
is based on the satellite line of sight information, which is derived
from the asteroseismology channel (which lies next to the exoplanet
channel on the focal plane). Currently, the pipeline only applies a
relative jitter correction for the three colour channels. This
correction conserves the total (white) flux, and no jitter correction
for the total flux is attempted. The pipeline also flags data points
collected during the SAA or affected by other events likely to impair
the data quality, such as entrance into and exit from the Earth's
shadow.

\onlfig{1}{
\begin{figure}
  \centering
  \epsfig{file=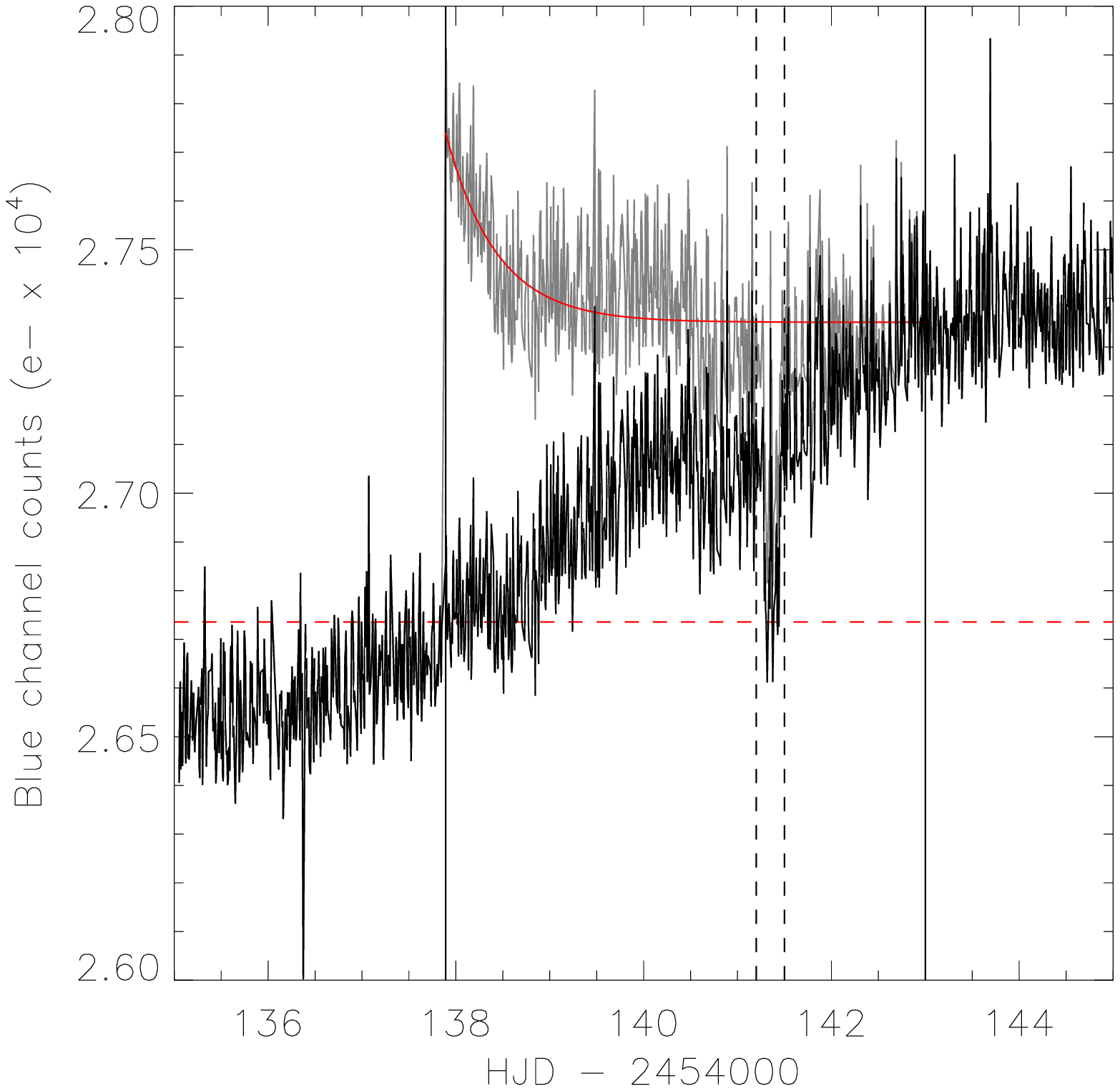, width=0.8\linewidth}
  \caption{Correction of the hot pixel event in the blue channel. The
    light curve before correction is shown in grey, and after
    correction in black. The vertical solid lines show the time limits
    of the correction section, with the exponential decay fit shown as
    the smooth red line. A small subsection, between vertical dashed
    lines, falls in-transit, and was not used to estimate the
    correction, though the correction was applied to it. The
    horizontal dashed line shows the reference flux level adopted
    before the hot pixel event.}
  \label{hotpix}
\end{figure}
}

The blue channel light curve is affected by a hot pixel event a few
days after the start of the observations, causing a sudden rise in the
measured flux followed by a gradual decay of $\sim 5$ days. This was
corrected by fitting an exponential curve to the decaying segment,
excluding a small portion that falls in-transit (see
Figure~\ref{hotpix}, online only). That section of the blue channel
light curve also shows a gradual rise in flux. To preserve it, a
linear rise between the local mean flux levels before and after the
section of light curve affected by the hot pixel was added after
subtracting the exponential decay. The red and green channels are free
of visible hot pixel events. Note that the transit depth is the same
in all three channels (within the uncertainties), as expected for a
planetary transit.

A single band-pass is sufficient for the present work, so the flux
from the three colour channels was summed to give a `white' light
curve (approximately covering the range $300$--$1000$\,nm). The
three-colour photometry will be discussed in an upcoming paper pending
improvements in the pipeline. Additionally, a version of the light
curve with regular 512\,s time sampling was computed by rebinning
there over-sampled part of the original. This version was used to
study the out-of-transit variability, while the over-sampled version
was retained to estimate the transit parameters. A short-baseline (5
data points) iterative non-linear filter \citep{ai04} with 5-$\sigma$
clipping was applied to both the over-sampled and regularly sampled
light curves to identify and reject further outliers, resulting in a
final duty cycle of 87\%.

\begin{figure}
  \centering
  \epsfig{file=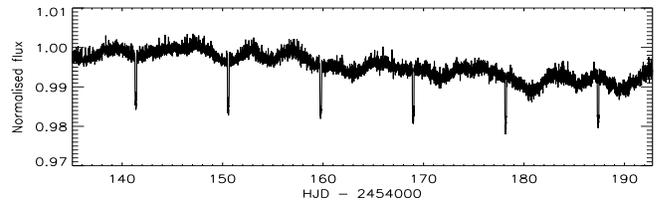, width=\linewidth, height=3cm}
  \caption{Pre-processed white light curve of CoRoT-Exo-4. The light
    curve was normalised by dividing it by its own median, which gives
    more weight to the later (oversampled) part of the light
    curve. This does not affect the transit analysis, since it is the
    local out-of-transit flux around each transit that is taken as a
    reference (see Section~\ref{transit}).}
  \label{fulllc}
\end{figure}

Based on preliminary ground-based imaging of the field \citep{dmd+06}
and a basic empirical model of the CoRoT PSF, we estimate the
contamination of the photometric aperture by other stars than
CoRoT-Exo-4 to be $0.3 \pm 0.1$\% (the uncertainty arises from the PSF
model). This estimate is compatible with the transit depths measured
from the ground (see Paper V). We therefore subtracted a
constant equal to 0.3\% of the median flux value, before normalising
the light curve (the uncertainty is accounted for separately, see
Section~\ref{transit}). The full normalised light curve is shown in
Figure~\ref{fulllc}. We evaluate the actual noise level per 512\,s by
measuring the dispersion about 1\,h (7 exposures) bins and scaling it
by $\sqrt{7}$, giving $8.9 \times 10^{-4}$, compared to a photon noise
level of $5.6\times 10^{-4}$. Possible factors contributing to the
difference include residual instrumental effects and the intrinsic
variability of the star.

\section{Transit  analysis}
\label{transit}

\begin{table}
  \caption[]{Star and transit parameters derived from the CoRoT light curve. 
     The quantity $M_{\rm
      s}^{1/3}/R_{\rm s}$, which is used in determining the stellar
    parameters (see Paper V), is derived directly from $a/R_{\rm S}$ and $P$.
  }
  \label{lcpar}
  $$
  \begin{array}{llc}
    \hline
    \noalign{\smallskip}
    \mathrm{Parameter} & \mathrm{Value} & \mathrm{Bayesian~range} \\
    \noalign{\smallskip}
    \hline
    \noalign{\smallskip}
    {\rm RA} & ~06~48~46.70 & \\
    {\rm Dec}& -00~40~21.97 & \\
    R-{\rm mag} & 13.45 & \\
    P_{\rm rot}~({\rm d}) & 8.87 \pm 1.12 & \\
    \noalign{\smallskip}
    \hline
    \noalign{\smallskip}
    P~({\rm d}) & 9.20205 \pm 0.00037 & \\
    T_0~({\rm HJD}) & 2454141.36416 \pm 0.00089 & \\
    \noalign{\smallskip}
    i~(^{\circ}) & 90.000_{-0.085}^{+0.000} & 87.708-90.000 \\
    \noalign{\smallskip}
    a/R_{\rm s} & 17.36_{-0.25}^{+0.05} & 14.30-17.80 \\
    \noalign{\smallskip}
    u & 0.44_{-0.15}^{+0.16} & 0.00-1.00 \\
    \noalign{\smallskip}
    R_{\rm p}/R_{\rm s} & 0.1047_{-0.0022}^{+0.0041} & 0.1000-0.1125 \\
    \noalign{\smallskip}
    M_{\rm s}^{1/3}/R_{\rm s} & 0.899_{-0.013}^{+0.003} & 0.741-0.922 \\
    \noalign{\smallskip}
    \hline
  \end{array}
  $$ 
\end{table}

A preliminary ephemeris (orbital period $P$ and epoch $T_0$) was
obtained by least-squares fitting of periodic, trapezoidal transits to
the light curve after filtering out the out-of-transit variations
using a 1-day baseline iterative non-linear filter. A more careful
removal of the variability was then carried out by fitting a straight
line to a light curve section lasting a little over one transit
duration before and after each transit. We also experimented with
higher order polynomials, but they did not improve the dispersion of
the residuals, and do not change the results of the subsequent
analysis. We folded the corrected segments of light curve using the
preliminary period ephemeris, rebinned them in bins of 0.0003 in
phase, and fitted the result to obtain preliminary estimates of the
system scale $a/R_{\rm s}$, the radius ratio $R_{\rm p}/R_{\rm s}$,
the inclination $i$ and the linear limb-darkening coefficient $u$,
where $R_{\rm s}$ is the star radius, $R_{\rm p}$ the planet radius
and $a$ the semi-major axis. We opted to fit for $u$ rather than
  fix it because reliable theoretical limb-darkening coefficients are
  not currently available for the CoRoT bandpass.  The ephemeris was
then refined by fitting for the time of transit centre $T_{\rm C}$ for
each individual transit event (fixing all other parameters) and
fitting a linear relation to the $T_C$'s. Finally, the light curve was
folded again at the refined ephemeris and rebinned to perform a final
fit for $a/R_{\rm s}$, $R_{\rm p}/R_{\rm s}$, $i$ and $u$.

At each stage, we use the formalism of \citet{ma02} with quadratic
limb darkening to generate model transit light curves and the {\sc
  Idl} implementation {\sc Mpfit} of the Levenberg-Marquart fitting
algorithm, kindly provided by C.\ Markwart, to perform the fit. The
period was fixed at the ephemeris value, and the epoch was also fixed
except when fitting individual transits. The eccentricity was assumed
to be zero (the best fit to the radial velocity data is a circular
orbit with an eccentricity uncertainty of 0.1, see Paper V). We also
tried fitting the transits with a quadratic limb-darkening
prescription, but this did not improve the fit, and therefore we
reverted to linear limb-darkening. 

\begin{figure}
  \centering \epsfig{file=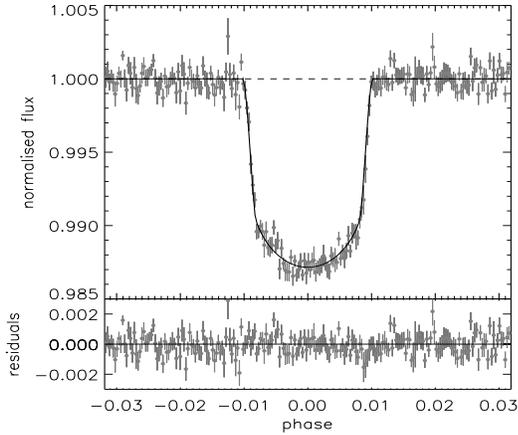, width=0.8\linewidth, height=6cm}
  \caption{Folded, binned light curve with the best-fit transit model.}
  \label{trfit}
\end{figure}

To evaluate the noise-induced uncertainties on the transit parameters,
including the effect of red noise, we used a `correlated bootstrap'
approach. The residuals from the global best fit were divided into
bins lasting 1.12\,h ($2/3$ of the transit duration, or $1/4$ of the
duration of the light curve segments used to calibrate the
out-of-transit variations around each transit), randomly shuffled, and
added back to the fit before fitting the individual
transits. Over-sampled and non over-sampled bins were shuffled
separately. Each bin is shuffled whole, preserving the detailed time
sampling of individual bins, so the procedure does not account for the
effect of small data gaps, but it does account for correlated noise on
hour timescales, including the effect of star spots crossed by the
planet. We used 100 realisations when fitting individual transits and
1000 when fitting the folded light curve. At each realisation, we also
added a constant drawn from a Gaussian distribution with zero mean and
standard deviation 0.001 to the data, to account for the uncertainty
in the contamination fraction. As the frequency distributions for each
parameter can be strongly non-Gaussian (see in Fig.~\ref{proba} in the
online material), we measure uncertainties as the interval away from
the best-fit value where the frequency drops below $e^{-1/2}$ times
the maximum (if the distributions were Gaussian, this would be
equivalent to the standard deviation). The results are reported in
Table~\ref{lcpar}. For comparison, we also show in Fig.~\ref{proba}
the results of a standard bootstrap (which simply consists in swapping
data points, and accounts for white noise only). Except for the
limb-darkening coefficient, the two processes give similar results,
indicating that red noise affects the other parameters' only slightly. 

\onlfig{4}{
\begin{figure*}
  \centering \epsfig{file=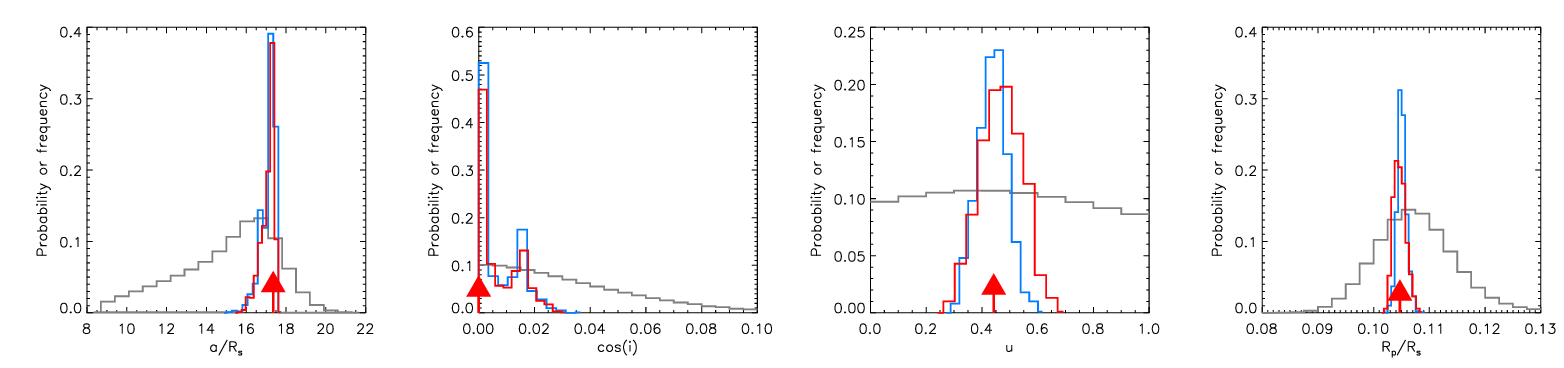, width=0.8\linewidth}
\epsfig{file=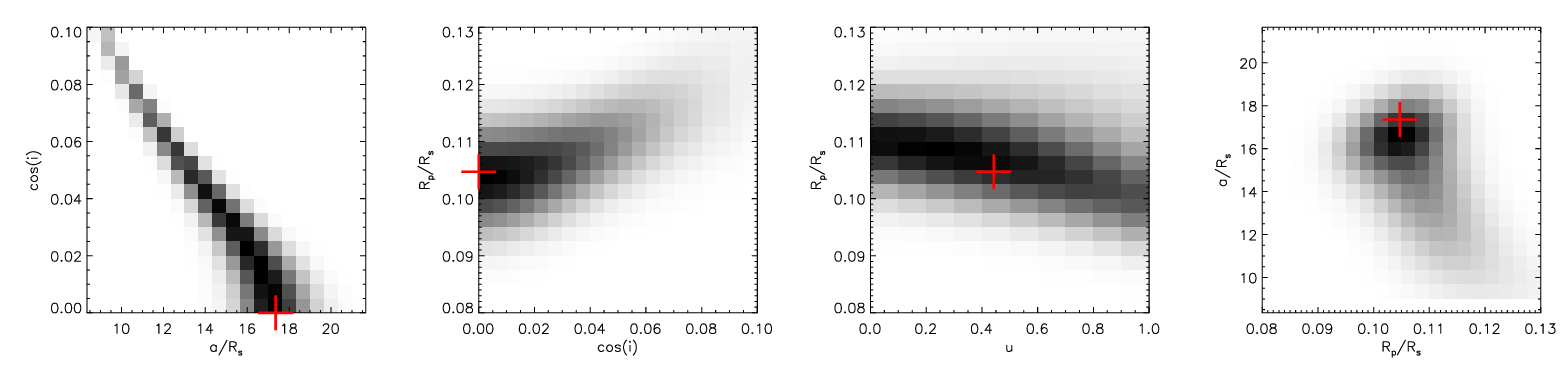, width=0.8\linewidth}
\caption{Top: probability and frequency distributions obtained from
  the Bayesian and bootstrap methods respectively for the main
  parameters, $a/R_{\rm s}$, $cos(i)$, $u$ and $R_{\rm p}/R_{\rm s}$
  of the transit fit. The Bayesian method results are shown in grey,
  the correlated bootstrap in red and the non-correlated bootstrap in
  blue. Bottom: Two-dimensional probability distributions obtained
  from the Bayesian method for selected pairs of parameters,
  highlighting the correlations between parameters (particularly
  $a/R_{\rm s}$ and $i$). In each panel, the location of the best-fit
  model is marked by a red vertical arrow or cross.}
  \label{proba}
\end{figure*}
}

To gain an insight into the effect of parameter-to-parameter
correlations, we used a Bayesian approach. Reduced $\chi^2$ values
were computed over a grid in 4-dimensional parameter space ($i$,
$a/R_{\rm s}$, $R_{\rm p}/R_{\rm s}$ and $u$) about the best fit. A
uniform grid in $\cos(i)$ was used, which is equivalent to assuming an
isotropic distribution of inclinations. The grid was uniform in the
other parameters, i.e.\ no a priori information on these parameters
was assumed (though $u$ was restricted to the physical range
0--1). The $\chi^2$'s were then converted to relative probabilities
for each individual model using $p \propto \exp(-\chi^2/2)$ and
normalised. One can then obtain probability distributions for each
parameter (shown on Fig.~\ref{proba}) by marginalising over successive
parameters. For each parameter, we report in Table~\ref{lcpar} the
interval over which the probability is higher than $e^{-1/2}$ times
the maximum. This interval should be interpreted with some care: it is
not a confidence interval in the frequentist sense, but rather an
interval containing $\sim 68\%$ of the posterior probability
\emph{integrated over all other parameters}\footnote{For an excellent
  discussion of Bayesian inference and model selection and how it
  differs from frequentist methods, see \protect\cite{tro08}.}. In the
present case, this interval is very wide because the global minimum in
the multi-dimensional $\chi^2$ surface is very narrow, but it is
located at one end of a valley which widens significantly away from
the minimum, as illustrated in Fig.~\ref{proba}.

In Paper V, the transit parameters are combined with ground-based
follow-up observations to give the stellar and planetary parameters,
which we reproduce here for completeness: $T_{\rm eff} = 6190 \pm
60$\,K, $\log g = 4.41 \pm 0.05$, $M_{\rm s} =
1.16_{-0.02}^{+0.03}\,M_{\odot}$, $R_{\rm s} =
1.17_{-0.03}^{+0.01}\,R_{\odot}$, age $1_{-0.3}^{+1.0}$\,Gyr; $M_{\rm
  p} = 0.72 \pm 0.08\,M_{\rm Jup}$ and $R_{\rm p} =
1.19_{-0.05}^{+006}\,R_{\rm Jup}$. Note that \citet{cla04}
  compute theoretical linear limb-darkening coefficients in the range
  0.55--0.6 for $T_{\rm eff}=6250$\,K and $\log g=4.5$ in $r'$, which
  is the standard bandpass closest to the peak of the CoRoT bandpass,
  though the later is much broader (300--1000\,nm). This is compatible
  with the value of $0.44 \pm 0.15$ obtained from the transit fit.

\onlfig{5}{
\begin{figure}
  \centering \epsfig{file=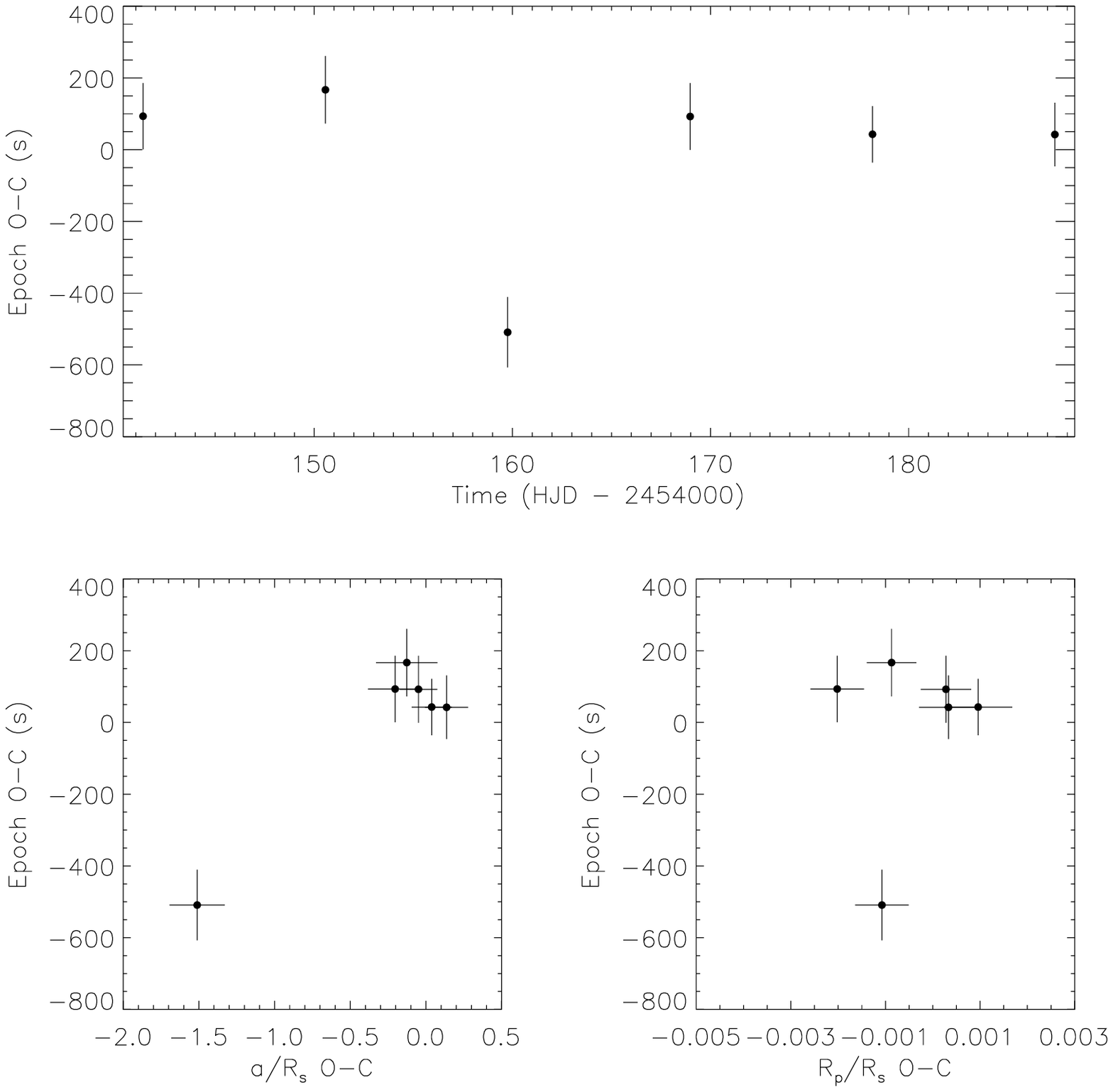, width=\linewidth}
  \caption{Transit timing residuals versus time (top panel) and
    compared to variations in transit duration (measured through the
    scale parameter $a/R_{\rm s}$) and depth (measured through the
    radius ratio $R_{\rm p}/R_{\rm s}$. The apparent timing deviation
    of the third transit is accompanied by a change in duration, but
    not in depth. Closer inspection of the light curve shows that this
    transit is more strongly affected by data gaps than the others.}
  \label{timing}
\end{figure}
}

We checked for transit timing variations in the $O-C$ (observed minus
computed) residuals from the refined ephemeris. The results are given
in Figure~\ref{timing} (online only). The third transit in the time
series shows a strong ($>500$\,s) deviation. To test whether this is a
real timing variation, we repeated the individual transit fits
allowing $a/R_{\rm s}$ and $R_{\rm p}/R_{\rm s}$ to vary as well as
$T_{\rm C}$, and found a clear correlations between the timing
residuals and $a/R_{\rm s}$ (see bottom left panel of
Figure~\ref{timing}), which points towards the effect of star spots or
instrumental systematics rather than a real timing variation as the
cause of the outlier. Closer inspection reveals that this transit
contains small data gaps, and we interpret the deviation in measured
timing and duration as an artefact of these gaps rather than a
physical effect. We also checked for a secondary eclipse (removing the
variability around phase 0.5 using linear fits as done for the
transits) but none is detected (as expected for this relatively low
irradiation planet).

\section{Stellar rotation}
\label{star}

\begin{figure}
  \centering
  \epsfig{file=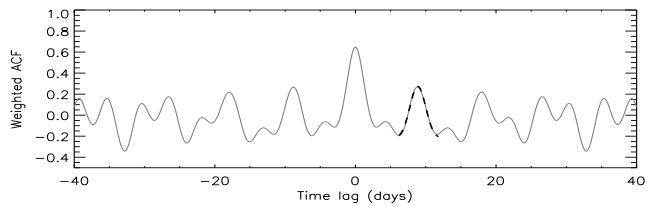, width=\linewidth, height=3cm}
  \caption{Weighted auto-correlation function of the out-of-transit
    light curve (gray line). We estimate the rotation period of the star using a
    Gaussian fit to the peak of the first side-lobe (dashed black line).}
  \label{acf}
\end{figure}

The light curve in Fig.~\ref{fulllc} shows clear out-of-transit
variability typical of a rotating, spotted photosphere. The
semi-coherent nature of the light curve suggests that active regions
on the star are evolving on timescales slightly longer than the
rotation period.  A given active region may therefore cross the
visible hemisphere only 2 or 3 times in its lifetime. The form of the
light curve evolves quickly, but partial coherency should persist for
2 or 3 rotations, making auto-correlation an appropriate method to
measure the stellar rotation period. We used an inverse variance
weighted adaptation of the Discrete Correlation Function method of
\citet{ek88} \citep{cds+08} to compute the Auto-Correlation Function,
or ACF (see Fig.~\ref{acf}). We fit a Gaussian to the first peak in
the ACF to deduce a period of $P_{\rm ACF}=8.87 \pm 1.12$\,d. The
dominant source of uncertainty is the limited duration of the
time-series. This is consistent with the $v \sin i$ of
$6.4 \pm 1.0$\,km/s reported in Paper V.

\section{Conclusions}
\label{concl}

CoRoT-Exo-4b's orbital period is the second longest of any transiting
planet to date, following HD\,17156b \citep{bal+07}. To our knowledge,
it is only the second transiting system where the star's photometric
rotation period has been measured photometrically (the other being
HD\,189733, \citealt{hw08}). There are other systems where the
rotation period and planet orbital period are very similar,
particularly $\tau$~Boo \citep{cds+07,dmf+08} and XO-3
\citep{jmb+08,wht+08} (the rotation periods are derived from Doppler
imaging and $v \sin i$ respectively, rather than measured
photometrically). In both cases, the period of the orbit and the
stellar rotation period are not exactly the same, but is compatible
with a synchronised outer envelope when allowing for differential
surface rotation. Both planets, which orbit F-type stars like
CoRoT-Exo-4, are massive and have very short periods ($\sim 3$\,d),
and thus may induce strong enough tides in the star to have
synchronised its convective envelope.

A simple calculation, based on the work of \citet{dlm04} and
\citet{jgb08}, shows that CoRoT-Exo-4b, which has a smaller mass and
longer period, would not exert a significant tidal torque on its host
star unless one invokes an unphysically high rate of tidal energy
dissipation in the star. However, \citet{mbv+08} discuss the case of
two other transiting systems containing F-type stars, XO-4 and
HAT-P-6, both with orbital periods longer than 7 days which are
roughly half the (spectroscopically determined) stellar rotation
period. They argue that this may indicate the existence of resonant
interactions between the planet's orbit and its rotating host star,
the absence of such resonances in systems containing cooler stars
suggesting that Jupiter-mass planets can only interact effectively
with stars with very shallow outer convective zones.

Similarly, if some other factor brought the CoRoT-Exo-4 system to the
1:1 ratio between stellar rotation and orbital period that is observed
today, resonant interaction between star and planet may have
maintained this ratio thereafter. Any subsequent evolution of the
star's rotation rate is likely to have been modest: 9\,d is close to
the peak of the rotation period distribution for F-type stars both
during the TTauri phase and in the field \citep[see e.g. Fig.~1
of][]{bar03}. The initial resonance may have occured naturally if the
proto-planetary disk was truncated near the co-rotation radius, and
the planets' migration halted close to the inner edge of the
disk. This hypothesis is supported by the absence of detectable
eccentricity (see Paper V).

This system clearly warrants further observational and theoretical
investigation to pin down its tidal and rotational evolution
status. For example, more detailed analysis of the out-of-transit
light curve should enable the active regions on the stellar surface to
be mapped in a time-resolved fashion \citep{lbr07} to search for signs
of star-planet magnetic interaction.

\begin{acknowledgements}
  HD and JMA acknowledge support from grant ESP2007-65480-C02-02 of
  the Spanish Science and Innovation ministry, the German CoRoT team
  (TLS and Univ.\ Cologne) from DLR grants 50OW0204, 50OW0603, and
  50QP0701, AL from contract ASI/INAF I/015/07/0 (work package 3170),
  and SZ from the Israel Science Foundation -- Adler Foundation for
  Space Research (grant no.\ 119/07).
\end{acknowledgements}

\bibliographystyle{aa} \bibliography{exo4b}

\end{document}